\documentclass[10pt]{article}
\usepackage{dcolumn}
\usepackage{bm}
\usepackage{cite}
\usepackage{amsthm}
\usepackage{amscd}
\usepackage{xspace}
\usepackage{bm}
\usepackage{amstext}
\usepackage{amsmath}
\usepackage{amsfonts}
\newcommand{\p}{\partial}

\newcommand{\dd}{{\rm d}}

\newcommand{\ed}{\end{definition}}                  
\newcommand{\bc}{\begin{corollary}}                 
\newcommand{\ec}{\end{corollary}}                   
\newcommand{\bl}{\begin{lemma}}                     
\newcommand{\el}{\end{lemma}}                       
\newcommand{\bp}{\begin{proposition}}            
\newcommand{\ep}{\end{proposition}}                
\newcommand{\bere}{\begin{remark}}                  
\newcommand{\ere}{\end{remark}}                     

\newcommand{\bt}{\begin{theorem}}
\newcommand{\et}{\end{theorem}}

\newcommand{\be}{\begin{equation}}
\newcommand{\ee}{\end{equation}}

\newcommand{\bit}{\begin{itemize}}
\newcommand{\eit}{\end{itemize}}
\newtheorem{theorem}{Theorem}[section]
\newtheorem{corollary}[theorem]{Corollary}
\newtheorem{lemma}[theorem]{Lemma}
\newtheorem{proposition}[theorem]{Proposition}
\theoremstyle{definition}

\theoremstyle{remark}
\newtheorem{remark}[theorem]{Remark}

\begin{document}

\title{Rayleigh's dissipation function at work}

\author{E. Minguzzi\thanks{
Dipartimento di Matematica e Informatica ``U. Dini'', Universit\`a
degli Studi di Firenze, Via S. Marta 3,  I-50139 Firenze, Italy.
E-mail: ettore.minguzzi@unifi.it} }


\date{}
\maketitle

\begin{abstract}
\noindent
It is shown that the Rayleigh's dissipation function can be successfully applied in the solution of mechanical problems involving friction non-linear in the velocities. Through the study of surfaces at contact we arrive at a simple integral expression which gives directly the Rayleigh dissipation function in terms of generalized coordinates. In this way the solutions of  Lagrangian problems with friction are  reduced to  often elementary calculations of the kinetic energy, the potential energy and the Rayleigh dissipation function. Some examples of pedagogical interest are given.
\end{abstract}

\maketitle

\section{Introduction}
On June 12th, 1873 Lord Rayleigh presented the memoir ``Some general theorems relating to vibrations'' to the London Mathematical Society \cite{strutt71}. In the second section entitled ``The dissipation function'' (see also section 81 of his 1877 monumental work ``The theory of sound'' \cite{rayleigh45}) he recalls that the conservative forces can be included into a potential function $V$ so that Lagrange equations read
\begin{equation} \label{all}
\frac{\dd }{\dd t} \frac{\p T}{\p \dot q^k}-\frac{\p T}{\p q^k}+\frac{\p V}{\p q^k}=Q_{k}^{(nc)},
\end{equation}
 leaving us with just the non-conservative generalized forces $Q_{k}^{(nc)}$.
He continues
\begin{quote}
The principal object of the present section is to show that another
group of forces [so far included in $Q^{(nc)}_k$, n.a.] may be advantageously treated in a similar manner.
The forces referred to are those which vary in direct proportion to
the component velocities of the parts of the system. It is well known
that friction, and other sources of dissipation, may be usefully represented as following this law approximately; and even when the true
law is different, the principal features of the case will be brought out.
\end{quote}
In a slightly more general form his argument is as follows. First recall that the Lagrange equations are obtained decomposing the mechanical system into $n$ idealized point particles of mass $m_a$ and positions ${\mathbf r_a}$, $a=1,\cdots, n$. The holonomic constraint is expressed through the dependence ${\mathbf r}_a(t,q)$ where $q^k$, $k=1,\cdots, l$ are the generalized coordinates, where $l\le 3n$. The Lagrange equations are then a restatement of D'Alembert's principle of virtual works (summation over repeated non-particle indices is understood; here $b=1,\cdots, n$)
\begin{equation}
\sum_b \big(-m_b {\mathbf a}_b+{\mathbf F}^{(active)}_b \big) \cdot \delta {\mathbf r}_b=0, \qquad \delta {\mathbf r}_b=\frac{\p {\mathbf r}_b}{\p q^k}\, \delta q^k,
\end{equation}
where ${\mathbf F}^{(active)}_b$ might include forces coming from the constraints that violate the frictionless nature (smoothness) of the constraint (so, for instance, friction due to contact contributes to this term). In Equation (\ref{all}) the generalized force is
\begin{equation} \label{kdx}
Q_k^{(nc)}=\sum_b {\mathbf F}^{(nc)}_b\cdot \frac{\p {\mathbf r}_b}{\p q^k}.
\end{equation}
where ${\mathbf F}^{(nc)}_b$ is the non-conservative active force acting on particle $b$.
It will also be useful to recall that
\begin{equation}
{\mathbf v}_a(t,q,\dot q)=\frac{\p {\mathbf r_a}}{\p q^k} \,\dot q^k+\frac{\p {\mathbf r_a}}{\p t} ,
\end{equation}
so that
\begin{equation} \label{msp}
\frac{\p {\mathbf v}_a}{\p\dot q^k}=\frac{\p {\mathbf r}_a}{\p q^k}.
\end{equation}
 Rayleigh considers the hypothesis that on particle $a$ acts a (Euclidean) non-conservative force linear in the velocities, which reads in Cartesian components (here $i,j=1,2,3$)
\begin{equation}
F_{a i}=-K_{ij} v^{j}_{a}, \qquad
\end{equation}
where $K$ is a symmetric matrix which might depend on time and on the positions $\{ {\mathbf r}_a \}$. This force can be written as $-\nabla_{{\mathbf v}_{a}} R$
where
\begin{equation}
R= \frac{1}{2} \sum_a  K_{ij} v^{i}_{a} v^{j}_{a}
\end{equation}
is termed {\em dissipation function} and $K$ is the {\em dissipation matrix}.
Thus recalling Eqs. (\ref{kdx}) and (\ref{msp})
\begin{equation}
Q_k^{(nc)}= - \sum_a  \nabla_{\!{\mathbf v}_{a}} \!R \, \frac{\p {\mathbf v}_a}{\p\dot q^k}+Q_k^{(nc')}=-\frac{\p R}{\p \dot q^k}+Q_k^{(nc')}
\end{equation}
where $Q_k^{(nc')}$ includes all the generalized non-conservative forces not taken into account by the Rayleigh term.
The Lagrange equations become
\begin{equation} \label{app}
\frac{\dd }{\dd t} \frac{\p T}{\p \dot q^k}-\frac{\p T}{\p q^k}+\frac{\p V}{\p q^k}=-\frac{\p R}{\p \dot q^k}+Q_k^{(nc')} ,
\end{equation}
where $R(t,q,\dot q)$ is a quadratic polynomial in $\dot{q}$, which is homogeneous of second degree when  the constraints are independent of time.
It can be observed that, for constraints independent of time, the power lost due to Rayleigh's friction forces is
\begin{equation}
\frac{\p R}{\p \dot q^k} \,\dot q^k= 2R.
\end{equation}
Rayleigh then goes on to apply (\ref{app}) to the study of the small oscillations of a system to which damping has been added.

The content of this memoir section can be found, almost unaltered, in the few textbooks on mechanics that mention the Rayleigh dissipation function \cite{whittaker17,meirovitch70,greenwood77,takwale79,goldstein01}. Given the fact that Rayleigh only considered the case of viscous (Stokes, linear) friction, many references \cite{meirovitch70,bullo05,fasano06,gans13} define the  Rayleigh function directly as a homogeneous quadratic form in the generalized velocities which returns the friction force. Very few references \cite{bloch96,marsden99} consider general friction forces which admit a dissipation potential, however, when this happens it remains as a theoretical suggestion and it is not explained which friction forces admit such potential and how $R$ should be calculated in practice. The only exception is the book by Lurie \cite{lurie02}, recently translated into English, were general friction forces are considered.

Given these limitations, the Rayleigh dissipation function was considered as a kind of mathematical artifice, perhaps worth mentioning for its historical interest but with almost no application saved for the special exception of viscous forces. The widespread belief that this function is only useful for linear friction, so often held in research papers \cite{riewe96}, became one of the motivations for the development of different  methods, often variational in nature, which included  Lagrangians containing fractional derivatives \cite{riewe96}, coupled systems \cite{bateman31}, or modified Lagrangians \cite{levicivita95,denman66,ray79,kobe86}, most of these techniques being rather special and still adapted  to the linear case.

In this work, joining Lurie \cite{lurie02}, I wish to provide some results and examples which clarify the usefulness of the Rayleigh dissipation function for non-linear frictions, including the Coulomb friction. It will become clear that the Lagrange equations are mostly useful for their covariance, not for their variational origin. Recall, that (\ref{app}) are covariant in the sense that under change of coordinates $q'(t,q)$ the equations for $q'$ are obtained from those for $q$ through multiplication by the Jacobian of the transformation.
As a consequence, the Rayleigh modification is equally useful as it preserves covariance though it has no variational interpretation.

\section{Surfaces at contact}

The Lagrange equations are mostly useful in the study of dry friction in mechanical problems involving contact between surfaces. Let us consider a body $B$ moving over a flat surface $S_0$.
Let $S_1$ be a flat surface of $B$ at  contact with $S_0$. For simplicity we may identify $B$ with $S_1$ collapsing normally all its mass into this surface.

\begin{remark}
This operation is practically justified only when $B$ is slender. If its  height is too large compared to the base then in the subsequent analysis of wear by friction one would have to consider the way $B$ tilts over $S_0$ as it gets consumed by friction on the base $S_1$. Such effect will be ignored (for a related analysis see \cite{villaggio01}). Observe that a full treatment of this problem would require the introduction of two  angular coordinates.
\end{remark}

Let us first consider  the simple case of uniform translational motion of velocity $v$ of $S_1$ over $S_0$, where $S_0$ is at rest. Let $N$ be the normal force between the surfaces, which we assume to be homogeneously distributed so that the pressure $p$ among the surfaces is independent of the point.
We shall assume that the dry friction between these surfaces depends only on the chemical nature of the surfaces, on the velocity $v$ and on the normal force $N$ between them as follows
\begin{equation}
{\mathbf F} =- N \mu(v) \hat {\mathbf v}
\end{equation}
where $\mu(v)$ is a dimensionless dynamical friction coefficient which depends on the module of the velocity. This function depends greatly on the surfaces at contact.
Typically, for non-lubricated friction $\mu(v)$ starts from some non-zero value, it decreases, and then it increases again with  velocity. For lubricated friction it starts linearly in the velocity, might decrease and reach a local minimum and then it increases again for large velocities, see \cite{braun11} and references therein. However, this behavior is not universal \cite{muser11}.  If $\mu$ is independent of velocity we have Coulomb friction while if $\mu$ is linear we have Stokes (viscous) friction.

As it stands this formula is not particularly useful since, in real life, the pressure between the surfaces will not be homogeneous and the velocity will depend on the relative motion which need not be translational as it might include some rotation.

Therefore, let us consider a more general case in which the flat surface $S_0$ is not necessarily  at rest, although we shall assume that it keeps its orientation (normal vector) on space. The movement of the flat surface $S_1$ over $S_0$ can be arbitrary although they have of course the same normal vector.

Let us decompose $S_1$ into $n$ regions of equal area $\Delta A$ which we idealize as point particles. The particle $a$, $a=1,\cdots, n$, has velocity ${\mathbf v}_a$ while the velocity of the point of $S_0$ instantaneously at contact with particle $a$ has velocity ${\mathbf b}_a$. Let
\begin{equation}
{\mathbf v}^{(r)}_a={\mathbf v}_a-{\mathbf b}_a
\end{equation}
be the relative velocity of $S_1$ with respect to $S_0$ at the $a$ location.
Then on particle $a$ acts a force
\begin{equation} \label{aas}
{\mathbf F}^{(nc)}_a=- p_a \,\Delta A\, \mu(v^{(r)}_a) \hat {\mathbf v}^{(r)}_a ,
\end{equation}
where $p_a$ is the pressure on the area element.
 Let $\int^v \!\!\mu \,\dd v$ be any indefinite integral of the function $\mu$. We define
\begin{equation}
R= \Delta A\,\sum_a p_a \int^{v^{(r)}_a} \!\!\! \!\!\mu \,\dd v,
\end{equation}
so that indeed
\begin{equation}
-\nabla_{\!{\mathbf v}_{a}} \!R= -\frac{\p R}{\p  v^{(r)}_a} \nabla_{\!{\mathbf v}_{a}} v^{(r)}_a=- \Delta A\, p_a\, \mu(v^{(r)}_a) \hat {\mathbf v}^{(r)}_a= {\mathbf F}^{(nc)}_a,
\end{equation}
thus
\begin{equation}
Q_k^{(nc)}= - \sum_a  \nabla_{\!{\mathbf v}_{a}} \!R \, \frac{\p {\mathbf v}_a}{\p\dot q^k}=-\frac{\p R}{\p \dot q^k}.
\end{equation}
Taking the limit $n\to \infty$, and denoting with $dA$ the area element, we obtain
\begin{equation}
R=\int_{S_1} \dd A \, p(x) \int^{v^{(r)}(x)} \!\!\!\!\!\!\!\!\mu \,\dd v.
\end{equation}
Thus the non-conservative generalized force can also be written
\begin{equation} \label{ave}
Q_k^{(nc)}= -N \overline{\mu(v^{(r)}) \frac{\p v^{(r)}}{\p \dot q^k}},
\end{equation}
where the average is made using the normalized pressure $p/N$ as weight.

Concerning the value of $p(x)$ we can make two possible assumptions. The simplest is that  the force $N$ is uniformly distributed hence $p$ is constant. However, for non-lubricated contact we can also apply Reye's assumption \cite{reye60,funaioli73} (see also  Holm's and Archard's work \cite{holm46,archard53}), which tells us that the pressure satisfies $p=k/v^{(r)}$, where $k$ is a normalization constant.

This inverse proportionality has the following explanation: we might imagine that at the microscopic level friction causes wear, and that the mass removed from $B$ on a certain region of the surface $S_1$ is proportional to the work done by friction forces on that region, that is, it is proportional to $p v^{(r)}$. But this work must be constant over regions with the same area, for otherwise as time passes the profile of the body $B$ would change, which would increase the pressure and hence the friction force precisely on those regions where we underestimated its action. Stated in another way, a constant product $pv^{(r)}$ is the only possibility that makes the profile of $B$ constant and hence the shape of $S_1$ stationary.

\begin{remark}
As a word of caution, this justification is somewhat  weak because, particularly for elastic materials, the shape of the surface at contact may change in dependence of the strength of the applied forces, and the assumption of stationarity for the profile could also fail. At the microscopic level  $S_1$ could be non-flat, for instance due to inhomogeneous stresses or because wear distributes between $S_0$ and $S_1$ in an inhomogeneous way rather than disappearing immediately from contact with the body $B$. On the contrary the argument requires the ideal flatness of $S_1$. Wear scarring the surface, wear hardening, and many other phenomena may alter the surface $S_1$ during friction modifying pressure in ways not accounted for by Reye's idealization.  Still Reye's assumption is quite elegant and allows us to illustrate  the dependence of the next results on a non-trivial function $p(v^{(r)})$.
\end{remark}


Clearly, the proportionality constant $k$ in Reye's hypothesis is such that
\begin{equation}
N=\int_{S_1} \frac{k}{v^{(r)}(x)} \, \dd A ,
\end{equation}
thus
\begin{align}
&\textrm{ {\em Homogeneous pressure:}} \nonumber\\
&R=N \big(\int_{S_1}  \dd A  \big)^{-1} \int_{S_1} \dd A  \int^{v^{(r)}(x)} \!\!\!\!\!\!\!\!\mu \,\dd v. \label{hom}\\
&\textrm{ {\em Reye's  hypothesis:}}  \nonumber\\
&R=N \big(\int_{S_1} \frac{\dd A}{v^{(r)}(x)} \,   \big)^{-1} \int_{S_1}  \, \frac{\dd A}{v^{(r)}(x)} \int^{v^{(r)}(x)} \!\!\!\!\!\!\!\!\mu \,\dd v. \label{rey}
\end{align}
Finally, we calculate the power lost due to friction forces. From Eq. (\ref{aas})
\[
P=-\sum_a {\mathbf F}^{(nc)}_a\cdot {\mathbf v}_a^{(r)}=\sum_a p_a \,\Delta A\, \mu(v^{(r)}_a)  { v}^{(r)}_a,
\]
which in the continuum limit $n\to \infty$ gives
\begin{align}
&\textrm{{\em Homogeneous  pressure:}} \nonumber \\
&P=N \big(\int_{S_1}  \dd A  \big)^{-1} \int_{S_1} \mu(v^{(r)}) \,  v^{(r)} \,\dd A. \\
&\textrm{ {\em Reye's  hypothesis:}} \nonumber \\
&P=N \big(\int_{S_1} \frac{\dd A  }{v^{(r)}(x)} \, \big)^{-1} \int_{S_1} \mu(v^{(r)}) \,\dd A.
\end{align}
This power is not necessarily $2R$ as for the original Rayleigh's viscous model.

The usefulness of Rayleigh's dissipation function for the solution of problems with (non-linear) friction should now be apparent. We have just to calculate $R$, and this can be done directly using generalized coordinates so that its expression will have a dependence on the generalized velocities. In other words we passed through a Cartesian study of the forces only to obtain the expression of the Rayleigh function, but once it has been established we should no more be involved with Cartesian coordinates. So, once $L$ and $R$  are calculated the equations of motion follow from Eq.\ (\ref{app}).

We feel that some examples may help to clarify the method.

\subsection{Exercise: The conveyor belt}

In 1860 Bouchet did some experiments to determine the velocity dependence of the dynamical friction coefficient $\mu$. One of the surfaces was made of iron while the other was made by wood, leather or iron \cite{masi97}. He found the approximate dependence
\[
\mu(v)=\frac{\mu_0-\mu_\infty}{1+a v}+\mu_\infty
\]
where $a,\mu_0,\mu_\infty$ are positive constants. Observe that the friction coefficient decreases till it reaches a constant value for high velocities. This dependence can probably be replaced by more precise results \cite{palmer49}, but  we shall assume its validity for the sake of the exercise. Coulomb friction is recovered for $a\to 0,\infty$.

So let us consider here a block of  wood over a conveyor belt. Let $(x,y)$ be horizontal Cartesian coordinates of the block on a frame at rest, and let the conveyor belt move at constant speed $(v_0,0)$. A person standing next to the belt, like in a grocery check-out line, pushes the block with a force $(F_x,F_y)$ and we assume for simplicity that the block does not rotate. We want to determine the equations of motion.

The result does not depend on whether we assume the Reye hypothesis or not since the relative velocity does not change with the point of contact.
The module of the relative velocity is
\[
v^{(r)}=\sqrt{(\dot x-v_0)^2+\dot y^2}.
\]
The Rayleigh's dissipation function is
\begin{align*}
R=N  \!\!\!\int^{v^{(r)}(x)} \!\!\!\!\!\!\!\!\!\!\!\!\!\!\!\mu(v) \,\dd v=N \big\{&\frac{\mu_0\!-\!\mu_\infty}{a}\ln \big[1\!\!+a \sqrt{(\dot x-v_0)^2\!+\dot y^2}\,\big]\\
&\quad +\mu_\infty  \sqrt{(\dot x-v_0)^2\!+\dot y^2} \,\big\}.
\end{align*}
Since $T=\frac{m}{2}(\dot x^2+\dot y^2)$ the equations of motion are
\begin{align*}
m\ddot x&=F_x\!-\!N \big\{\frac{\mu_0-\mu_\infty}{1\!\!+a \sqrt{(\dot x-\!v_0)^2+\dot y^2}} +\mu_\infty\big\} \frac{\dot x-v_0}{\sqrt{(\dot x-\!v_0)^2+\dot y^2}},\\
m\ddot y&=F_y\!-\!N \big\{\frac{\mu_0-\mu_\infty}{1\!\!+a \sqrt{(\dot x-\!v_0)^2+\dot y^2}} +\mu_\infty\big\} \frac{\dot y}{\sqrt{(\dot x-\!v_0)^2+\dot y^2}} ,
\end{align*}
which could have been obtained directly from Eq.\ (\ref{ave}).
Observe that if $F_x$ is used to keep the block at $x=0$ then it becomes easy to move the block transversally across the belt since for low speeds $\dot y$ the effective transversal friction is  linear in the velocity.

\subsection{Exercise: A rotating disk}

Let us consider a disk $D$ of mass $m$ and radius $r$ placed on horizontal surface. At time $t=0$ the  disk has angular  velocity $\omega$ about its center and zero translational velocity. We wish to find at which time $\tau$ it will stop under Coulomb friction and Reye's assumption.

Let $(\rho, \theta)$ be polar coordinates on the disk and let $\varphi$ be our generalized coordinate, namely the angle of rotation of the disk with respect to a reference abscissa. The velocity at $P(\rho,\theta)$ is $v=\dot \varphi \rho$.
The Lagrangian $L=T=\frac{mr^2}{4}\, \dot \varphi^2$  so we have just to calculate $R$ which according to Eq.\ (\ref{rey}) is
\[
R=\mu mg [\int_{D} \frac{1}{\dot \varphi \rho} \, \dd \theta \rho \dd \rho ]^{-1} \int_{D} \dd \theta \rho \dd \rho = \frac{\mu mgr}{2}\, \dot \varphi .
\]
The Lagrange equation is
\[
\frac{mr^2}{2} \,\ddot \varphi=-\frac{\mu mgr}{2}
\]
thus the disk stops after a time interval $\tau=\frac{\omega r}{\mu g }$.

\subsection{Exercise: The rotating stone polisher}
Let us consider a rotating stone polisher which we describe as two concentric counter-rotating rings of approximately the same mass $m$ and radius $r$ placed on a horizontal rough surface  (in reality the rings are actually disks and they are not concentric but we wish to simplify the model). A worker handles the polisher imparting a force ${\mathbf F}=(F_x,F_y)$ and a momentum $M$ to it.
Let $(x,y)$ denote the position of the center and let $\theta$ be the orientation of the machine. We assume Coulomb friction and that the pressure over the rings is uniformly distributed. Let $\omega$ be the angular velocity of the rings, this means that the rings are  rotated of an angle $\theta\pm \omega t$ with respect to a reference abscissa. If the  angle $\varphi$ is used to label the points along the ring, the  velocity of the generic point is
\[
{\mathbf v}=(\dot x-r\sin(\theta\pm \omega t+\varphi)(\dot \theta\pm\omega), \dot y+r\cos(\theta\pm \omega t+\varphi)(\dot \theta\pm \omega)),
\]
thus
\begin{align*}
v_{\pm}^2&=\dot x^2+\dot y^2+r^2(\dot \theta\pm \omega)^2\\
&\quad +2r(\dot \theta\pm \omega)[-\dot x\sin(\theta\pm \omega t+\varphi)+\dot y \cos(\theta\pm \omega t+\varphi)]
\end{align*}
Let $(u, \phi)$ be the velocity of the center of the ring in polar coordinates on the velocity space. We have
\[
v_{\pm}^2=u^2+r^2(\dot \theta\pm \omega)^2-2 u r(\dot \theta\pm \omega)\sin(\theta\pm \omega t+\varphi-\phi).
\]
We are interested only in the range $r\omega \gg u, r \dot\theta$ which means that the polisher is working and rotating sufficiently fast with respect to the translational and rotational velocity imparted by the agent worker. In this limit
\begin{align*}
v_{\pm}&=r  \omega[1\pm
\frac{\dot \theta}{\omega} \mp \frac{u}{r \omega}\sin(\theta\pm \omega t+\varphi-\phi)\\
&\quad +\frac{1}{2} (\frac{u}{r \omega})^2 (1-\sin^2(\theta\pm \omega t+\varphi-\phi))]
\end{align*}
up to quadratic terms in the velocity ratios.

Using Eq.\ (\ref{hom}) the function $R$ is $R=R_++R_-$ where
\begin{align*}
R_{\pm}&=\frac{\mu m g}{2\pi} \int_0^{2\pi} \!\!\!v_\pm(\varphi) \,\dd \varphi
\simeq\mu m g\big(r \omega\pm r \dot \theta+\frac{1}{4} \frac{\dot x^2+\dot y^2}{r \omega}\big)
\end{align*}
Observe that the Rayleigh's function is expressed directly in terms of the generalized coordinates $(x,y,\theta)$ and the generalized velocities. Since
\[
T=\frac{1}{2}\, m [2(\dot x^2+\dot y^2)+2 r^2(\dot \theta^2+\omega^2)]
\]
the dynamical equations are
\begin{align*}
2m \ddot x&=-\mu m g \frac{\dot{x}}{r \omega} +F_x,\\
2m \ddot y&=-\mu m g \frac{\dot{y}}{r \omega} +F_y,\\
2 m r^2\ddot \theta &= M.
\end{align*}
We recover the  known and remarkable fact that when the polisher is active and rotating the Coulomb friction on its parts acts, as a whole, as a Stokes friction for the translational degrees of freedom, while for the rotating degree of freedom at the lowest orders the friction does not introduce neither a constant (Coulomb) nor a linear (Stokes) component. Thus moving the polisher  is easy when it is on and becomes difficult when it is off.

\section{Conclusions}

In textbooks and specialized articles it is widely held that the Rayleigh dissipation function is really useful only for  linear (Stokes, viscous) friction. As a consequence, it has been often regarded as a mathematical curiosity along with many other ad hoc methods that have been conceived along the years to deal with this case.
On the contrary, we have shown in this paper that the  Rayleigh dissipation function can be effectively used for non-linear friction as well, with apparently no  theoretical limitations.

In a sense a somewhat negative historical attitude towards the Rayleigh dissipation function, one which prevented a more widespread knowledge of its full potentialities, can be explained recalling the success of the variational methods in modern physics. This fact, together with the variational origin of the Lagrange equation has brought
many authors to the  identification of the Lagrangian methods with the variational methods. From here a kind of uneasiness with the non-variational nature of the Rayleigh modification to the Lagrange equations can be easily understood. However, even historically, the Lagrangian methods should not be identified with the variational methods as the Lagrange equations were developed from the principle of virtual works. Its main merit was that of providing {\em covariant} equations. Now, while it is true that the variation of an action leads to a covariant equation, not all covariant equations arise in this way. Once it is understood that covariance is much more at the heart of the success of Lagrangian mechanics than its variational origin it becomes easy to accept the Rayleigh dissipation function and the consequent treatment of friction phenomena as an equally worthy part of Lagrangian mechanics.

Of course, it is understood that the method applies only to  friction phenomena which can be phenomenologically described with just a response function $\mu(v^{(r)})$. In many physical applications this is not really the case and in fact in some circumstances friction might not be treatable within  the realm of classical mechanics \cite{urbakh04}.

As a last comment, we hope that the results of this work, by broadening the applicability of the Rayleigh dissipation function could prove useful to the scientist and to the teacher alike as these results might be used to generate many  interesting problems and exercises.

\section*{Acknowledgments}   I thank the referees for some useful suggestions. This work has been
partially supported by GNFM of INDAM.


\end{document}